\documentclass[aps,prl,10pt,floatfix,showpacs,superscriptaddress,twocolumn]{revtex4-1}
\usepackage[T1]{fontenc}
\usepackage{amsmath}
\usepackage{amssymb}
\usepackage[english]{babel}
\usepackage{booktabs}
\usepackage{graphicx}
\usepackage[colorlinks=true,citecolor=blue,linkcolor=blue]{hyperref}
\usepackage{multirow}
\usepackage{newtxmath}
\usepackage{newtxtext}

\newcommand{\Eq}[1]{Eq.~(\ref{#1})}
\newcommand{\e}{{\rm e}}
\newcommand{\s}{\mathbf{s}}

\begin{document}

\title{Solving Statistical Mechanics Using Variational Autoregressive Networks}

\author{Dian Wu}
\affiliation{School of Physics, Peking University, Beijing 100871, China}
\author{Lei Wang}
\email{wanglei@iphy.ac.cn}
\affiliation{Institute of Physics, Chinese Academy of Sciences, Beijing 100190, China}
\affiliation{CAS Center for Excellence in Topological Quantum Computation, \\ University of Chinese Academy of Sciences, Beijing 100190, China}
\affiliation{Songshan Lake Materials Laboratory, Dongguan, Guangdong 523808, China}
\author{Pan Zhang}
\email{panzhang@itp.ac.cn}
\affiliation{Key Laboratory of Theoretical Physics, Institute of Theoretical Physics, \\ Chinese Academy of Sciences, Beijing 100190, China}

\begin{abstract}
We propose a general framework for solving statistical mechanics of systems with finite size.
The approach extends the celebrated variational mean-field approaches using autoregressive neural networks, which support direct sampling and exact calculation of normalized probability of configurations.
It computes variational free energy, estimates physical quantities such as entropy, magnetizations and correlations, and generates uncorrelated samples all at once.
Training of the network employs the policy gradient approach in reinforcement learning, which unbiasedly estimates the gradient of variational parameters.
We apply our approach to several classic systems, including $2$D Ising models, the Hopfield model, the Sherrington--Kirkpatrick model, and the inverse Ising model, for demonstrating its advantages over existing variational mean-field methods. Our approach sheds light on solving statistical physics problems using modern deep generative neural networks.
\end{abstract}

\maketitle

Consider a statistical physics model such as the celebrated Ising model, the joint probability of spins $\s \in \{\pm 1\}^N$ follows the Boltzmann distribution
\begin{equation}
p(\s) = \frac{\e^{-\beta E(\s)}}{Z},
\label{eq:boltzmann}
\end{equation}
where $\beta = 1/T$ is the inverse temperature and $Z$ is the partition function. Given a problem instance, \emph{statistical mechanics} problems concern about how to estimate the free energy $F = -\frac{1}{\beta} \ln Z$ of the instance, how to compute macroscopic properties of the system such as magnetizations and correlations, and how to sample from the Boltzmann distribution efficiently. Solving these problems are not only relevant to physics, but also find broad applications in fields like Bayesian inference where the Boltzmann distribution naturally acts as posterior distribution, and in combinatorial optimizations where the task is equivalent to study zero temperature phase of a spin-glass model.

When the system has finite size, computing exactly the free energy belongs to the class of \#P-hard problems, hence is in general intractable. Therefore, usually one employs approximate algorithms such as variational approaches.
The variational approach adopts an ansatz for the joint distribution $q_\theta(\s)$ parametrized by variational parameters $\theta$, and adjusts them so that $q_\theta(\s)$ is as close as possible to the Boltzmann distribution $p(\s)$. The closeness between two distributions is measured by Kullback--Leibler (KL) divergence~\cite{mackay2003information}
\begin{equation}
D_\text{KL}(q_\theta \,\|\, p) = \sum_\s q_\theta(\s) \ln \left( \frac{q_\theta(\s)}{p(\s)} \right) = \beta (F_q - F),
\end{equation}
where
\begin{equation}
F_q = \frac{1}{\beta} \sum_\s q_\theta(\s) \left[ \beta E(\s) + \ln q_\theta(\s) \right]
\label{eq:vf}
\end{equation}
is the variational free energy corresponding to distribution $q_\theta(\s)$. Since the KL divergence is non-negative, minimizing the KL divergence is equivalent to minimizing the variational free energy $F_q$, an upper bound to the true free energy $F$.

One of the most popular variational approaches, namely the variational mean-field method, assumes a factorized variational distribution $q_\theta(\s) = \prod_i q_i(s_i)$, where $q_i(s_i)$ is the marginal probability of the $i$th spin. In such parametrization, the variational free energy $F_q$ can be expressed as an analytical function of parameters $q_i(s_i)$, as well as its derivative with respect to $q_i(s_i)$. By setting the derivatives to zero, one obtains a set of iterative equations, known as the \emph{na\"ive mean-field} (NMF) equations. Despite its simplicity, NMF has been used in various applications in statistical physics, statistical inference and machine learning~\cite{jordan1999, bishop2006pattern}. Although NMF gives an upper bound to the physical free energy $F$, typically it is not accurate, since it completely ignores the correlation between variables. Other approaches, which essentially adopt different variational ans\"atze for $q_\theta(\s)$, have been developed to give better estimate (although not always an upper bound) of the free energy. These ans\"atze, including Bethe approximation~\cite{bethe1935, yedidia2001}, Thouless--Anderson--Palmer equations~\cite{thouless1977solution}, and Kikuchi loop expansions~\cite{kikuchi1951theory}, form a family of mean-field approximations~\cite{jordan1999}.

However, on systems with strong interactions and on a factor graph with loops of different lengths (such as lattices), mean-field approximations usually give very limited performance. The major difficulty for the mean-field methods in this case is to give a powerful, yet tractable variation form of joint distribution $q_\theta(\s)$.
In this Letter, we generalize the existing variational mean-field methods to a much more powerful and general framework using autoregressive neural networks.

\paragraph{Variational autoregressive networks.}

The recently developed neural networks give us ideal methods for parameterizing variational distribution $q_\theta(\s)$ with a strong representational power. The key ingredient of employing them to solve statistical mechanics problem is to design neural networks such that the variational free energy \Eq{eq:vf} is efficiently computable. The method we adopted here is named \emph{autoregressive networks}, where the joint probability of all variables is expressed as product of conditional probabilities~\cite{dlbook, frey1998graphical, nade, made}
\begin{equation}
q_\theta(\s) = \prod_{i = 1}^N q_\theta(s_i \mid s_1, \ldots, s_{i - 1}),
\label{eq:autoregressive}
\end{equation}
and the factors are parametrized as neural networks. We denote using \Eq{eq:autoregressive} as an ansatz for the variational calculation of \Eq{eq:vf} as \emph{variational autoregressive networks} (VAN) approach for statistical mechanics.

\begin{figure}[tb]
\centering
\includegraphics[width=\linewidth]{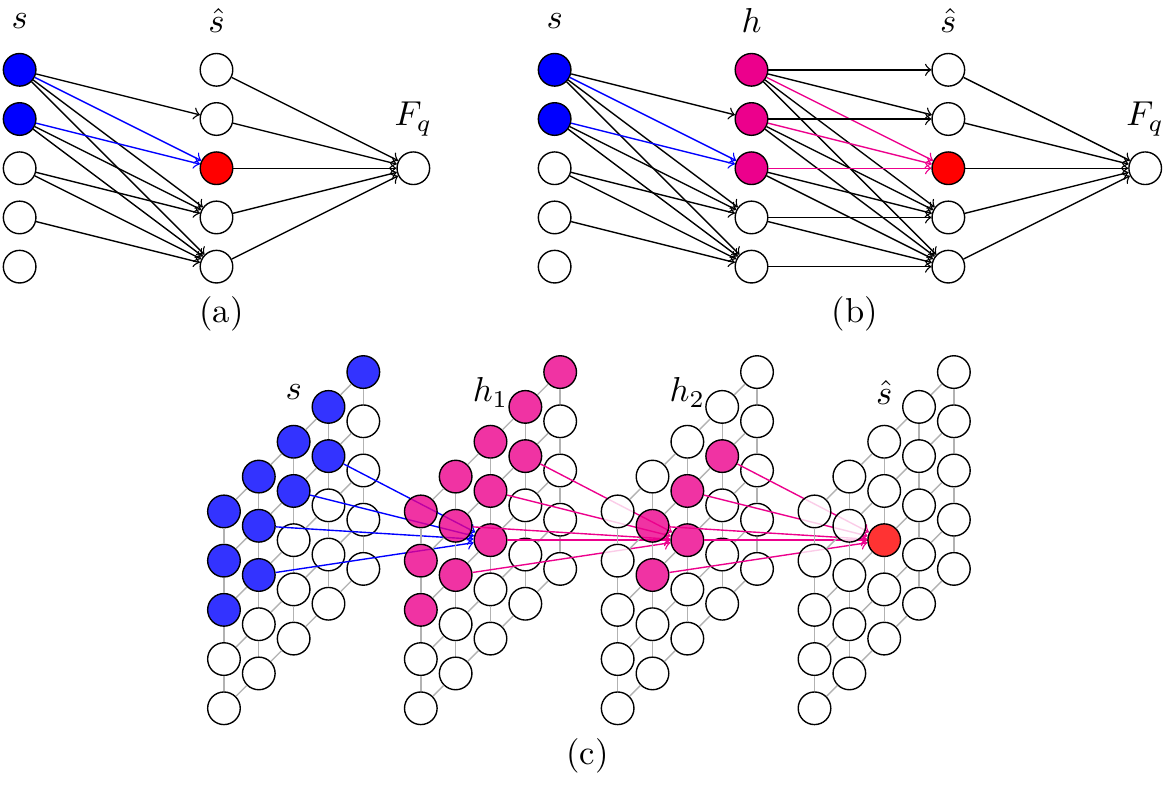}
\caption{Autoregressive networks with different architectures for variational free energy calculation. The spin configuration $\s$ is the input to the network, $\hat{\s}$ is the output of the network, and $h$ denotes hidden layer. The loss function $F_q$ is given by \Eq{eq:vf} and \Eq{eq:autoregressive}. The colored sites denote the receptive field of a site in $\hat{\s}$.
(a) The network has only one layer, which is densely connected, while the autoregressive property hold.
(b) The network has a hidden layer.
(c) The network has masked convolution layers on $2$D lattice. Only connections in a convolution kernel are shown for clarity.}
\label{fig:autoreg}
\end{figure}

The simplest autoregressive network is depicted in Fig.~\ref{fig:autoreg}(a), which is known as the \emph{fully visible sigmoid belief network}~\cite{frey1998graphical}. The input of the network is a configuration $\s \in \{\pm 1\}^N$ with a predetermined order, and the output
$\hat s_i = \sigma \left( \sum_{j < i} W_{i j} s_j \right)$
has the same dimension as the input. We see that the network is parametrized by a triangular matrix $W$, which ensures that $\hat s_i$ is independent with $s_j$ when $j \geq i$. This is named as \emph{autoregressive property} in machine learning literatures.
The sigmoid activation function $\sigma(\cdot)$ ranges in $(0, 1)$, so we can expect that $\hat s_i$ represents a probability with proper normalization. Namely, $\hat s_i = q(s_i = +1 \mid \s_{< i})$, which means the conditional probability of $s_i$ being $+1$, given the configuration of spins in front of it, $\s_{<i}$, in the predetermined order of variables.
Thus, given a configuration $\s$ as the input to the network, the joint distribution of the input variables can be expressed as the product of conditional probabilities, and each factor is a Bernoulli distribution $q(s_i \mid \s_{< i}) = \hat s_i \delta_{s_i, +1} + (1 - \hat s_i) \delta_{s_i, -1}$.

There have been many discussions in the machine learning community on how to make the autoregressive network deeper and more expressive, and how to increase the generalization power by sharing weights~\cite{bengio2000modeling, larochelle2011neural, gregor2013deep, nade, made}.
Using the simplest one-layer network as building blocks, we can design more complex and expressive networks, while preserving the autoregressive property. For example, we can add more layers of hidden variables to the network, as shown in Fig.~\ref{fig:autoreg}(b).

When the system has structures, e.g. lying on a $2$D lattice, a classic network architecture designed specifically for it is the convolutional network~\cite{dlbook}, which respects the locality and the translational symmetry of the system. To ensure the autoregressive property, one can put a mask on the convolution kernel, so that the weights are not zero only for half of the kernel, and $\hat s_i$ is independent of $s_j$ with $j < i$ in the predetermined order. The receptive field of the masked convolution through multiple layers is shown in Fig.~\ref{fig:autoreg}(c). This kind of structured autoregressive networks is known as \textsc{PixelCNN}~\cite{pixelcnn}, which has achieved state-of-the-art results in modeling and generating natural images. In additional, by using the dilated convolutions the autoregressive \textsc{WaveNet}~\cite{wavenet} can capture long-range correlations in audio signals, and has achieved remarkable performance in real-world speech synthesis.

The autoregressive networks are one of the leading generative models that find wide applications under the general purpose of density estimations~\cite{pixelcnn, wavenet, papamakarios}.
A key difference between our work and those machine learning applications is that for density estimation one trains the network from the training data using maximum likelihood estimation, i.e. minimizing the KL divergence between empirical training data distribution $p_\text{data}(\s)$ and the network, $D_\text{KL}(p_{\text{data}} \,\|\, q_\theta)$. Whereas in our variational free energy calculation, the goal is to reduce the \emph{reversed} KL divergence $D_\text{KL}(q_\theta \,\|\, p)$. Therefore, we train the network using data produced by itself. The only input of our calculation is the energy function of the statistical mechanics problem, and no training data from the target Boltzmann distribution is assumed.

The variational free energy in \Eq{eq:vf} can be regarded as a scalar loss function over the parameters $\theta$ of the autoregressive network of \Eq{eq:autoregressive}. A nice feature of autoregressive networks is that one can draw independent samples efficiently by sampling each variable in the predetermined order.
Moreover, one has direct access to the normalized probability $q_{\theta}(\s)$ of any given sample. Exploiting these properties, one can replace the summation over all possible configurations weighted by $q_\theta(\s)$ by samplings from the network, and evaluate the entropy and energy terms respectively in \Eq{eq:vf}. Thanks to the direct-sampling ability, the estimated variational free energy provides an exact upper bound to the true free energy of the model.

The gradient of the variational free energy with respect to network parameters reads~\cite{sm}
\begin{equation}
\beta \nabla_\theta F_q = \mathbb{E}_{\s \sim q_\theta(\s)} \left\{ \left[ \beta E(\s) + \ln q_\theta(\s) \right] \nabla_\theta \ln q_\theta(\s) \right\}.
\label{eq:gradient}
\end{equation}
We perform the stochastic gradient descent optimization on the parameters $\theta$. Furthermore, we employ the control variates method of Ref.~\cite{mnih2014neural} to reduce the variance in the gradient estimator~\cite{sm}.
In the context of reinforcement learning~\cite{sutton1998reinforcement}, $q_\theta(\s)$ is a stochastic policy which produces instances of $\s$, and the term in the square bracket of \Eq{eq:vf} is the reward signal. Thus, learning according to \Eq{eq:gradient} amounts to the policy gradient algorithm.
We note that the variational studies of quantum states~\cite{carleo2017solving} employ a similar gradient estimator. However, the variational autoregressive networks enjoy unbiased estimate of the gradient using efficient direct sampling instead of relying on the correlated Markov chains.

To the best of our knowledge, the variational framework using deep autoregressive networks for statistical mechanics has not been explored before. Our method can be seen as an extension to the variational mean-field methods with a more expressive variational ansatz. Its representational power comes from recently developed deep neural networks with guarantee of \emph{universal expressive power}~\cite{dlbook}.
Rather than a specific model, we consider our approach as a general framework, analogous to existing frameworks such as Markov chain Monte Carlo (MCMC), mean-field methods, and tensor networks~\cite{physrevlett.99.120601, physrevlett.103.160601}.
When compared with existing frameworks, the features of VAN are these: giving an upper bound to the true free energy; efficiently generating independent samples without needing Markov chains, which is ideal for parallelization (on GPUs); and computing physical observables, such as the energy and correlations, using a sufficiently large amount of samples without any autocorrelations.

\paragraph{Numerical experiments.}

To demonstrate the ability of VAN in terms of accuracy of the variational free energy and estimated physical quantities, we perform experiments on Ising models. The energy of the configuration $\s$ is given by
$E(\s) = -\sum_{(i j)} J_{i j} s_i s_j$,
with $(i j)$ denoting pair of connections.
With different choices of the coupling matrix $J$, we cover systems on different topologies: $2$D square and triangular lattices, and fully connected systems. We also cover systems with different behaviors: ferromagnetic, antiferromagnetic, glassy, and as associative memory.

We first apply our approach to the ferromagnetic Ising model on $2$D square lattice with periodic boundary condition, which admits an exact solution~\cite{onsager}.
We have tested two types of network architectures, the $2$D convolution (conv) and densely connected (dense) respectively, to verify that taking into account the lattice structure is beneficial. More details on the implementation are discussed in the Supplemental Material~\cite{sm}.

\begin{figure}[tb]
\centering
\includegraphics[width=0.48\linewidth]{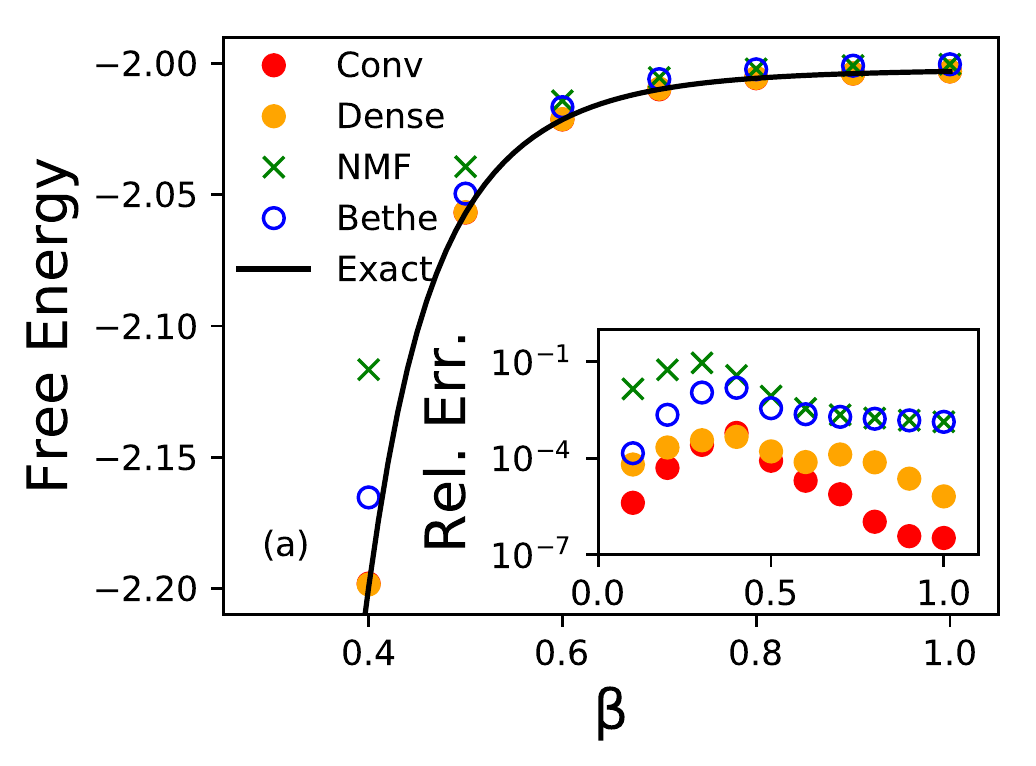}
\includegraphics[width=0.48\linewidth]{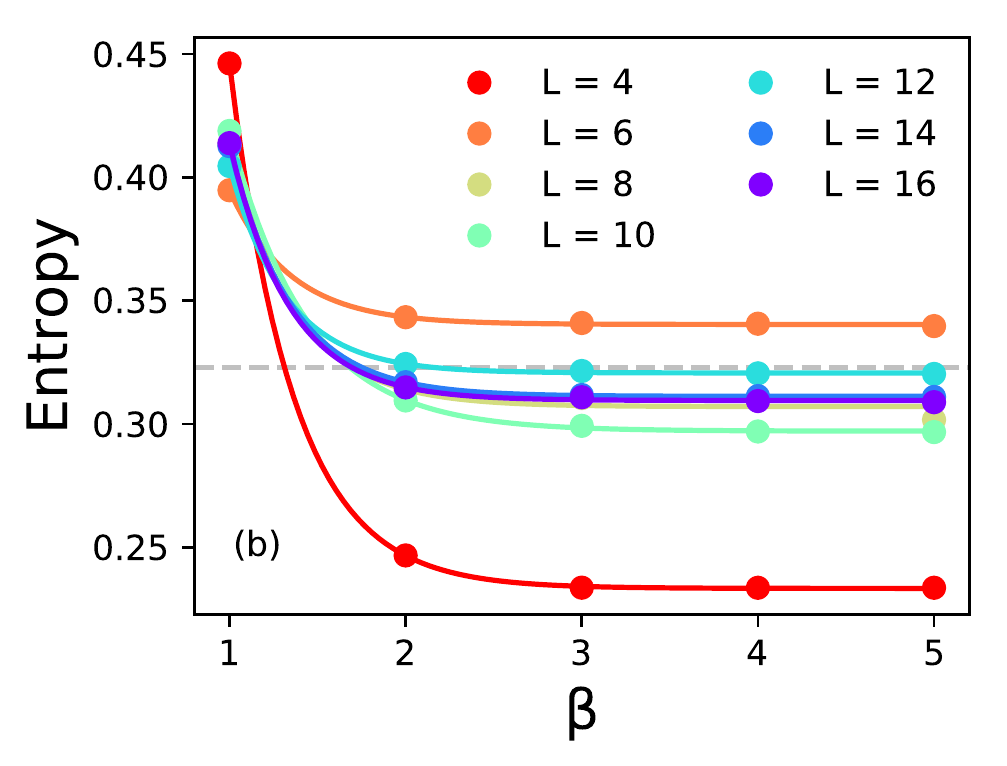}
\caption{(a) Free energy per site and its relative error of ferromagnetic Ising model on $16\times 16$ square lattice with periodic boundary condition.
(b) Entropy per site of antiferromagnetic Ising model on triangular lattices of various sizes $L$ with periodic boundary condition. The exact result (dashed line) at $T = 0$ and $L \to \infty$ is $S/N = 0.323066$~\cite{physrev.79.357, physrevb.7.5017}. The curves for $L = 8, 14, 16$ are almost overlapped.}
\label{fig:fm-ising}
\end{figure}

The free energy given by VAN, compared with NMF and Bethe approximations, is shown in Fig.~\ref{fig:fm-ising}(a). The figure shows that VAN significantly outperforms the two traditional methods. The maximum relative error is around the critical point, where the system develops long range correlations. Also, the network architecture with convolution layers performs significantly better than dense connection, since it respects the two-dimensional nature of the lattice, which is particularly beneficial when the correlation is short ranged. However, around criticality, they exhibits similar performance.

Then, we apply our approach to the frustrated antiferromagnetic Ising model on $2$D triangular lattice with a periodic boundary condition.
Fig.~\ref{fig:fm-ising}(b) shows the entropy per site versus inverse temperature $\beta$ for various lattice sizes. Reaching a finite entropy density indicates that the system processes an exponentially large number of degenerate ground states. Extrapolation of $\beta \to \infty$ shows that VAN correctly captures the exponentially large number of ground states. In comparison, describing such feature has been challenging to conventional MCMC and mean-field approaches.

\begin{figure}[tb]
\centering
\includegraphics[width=0.48\linewidth]{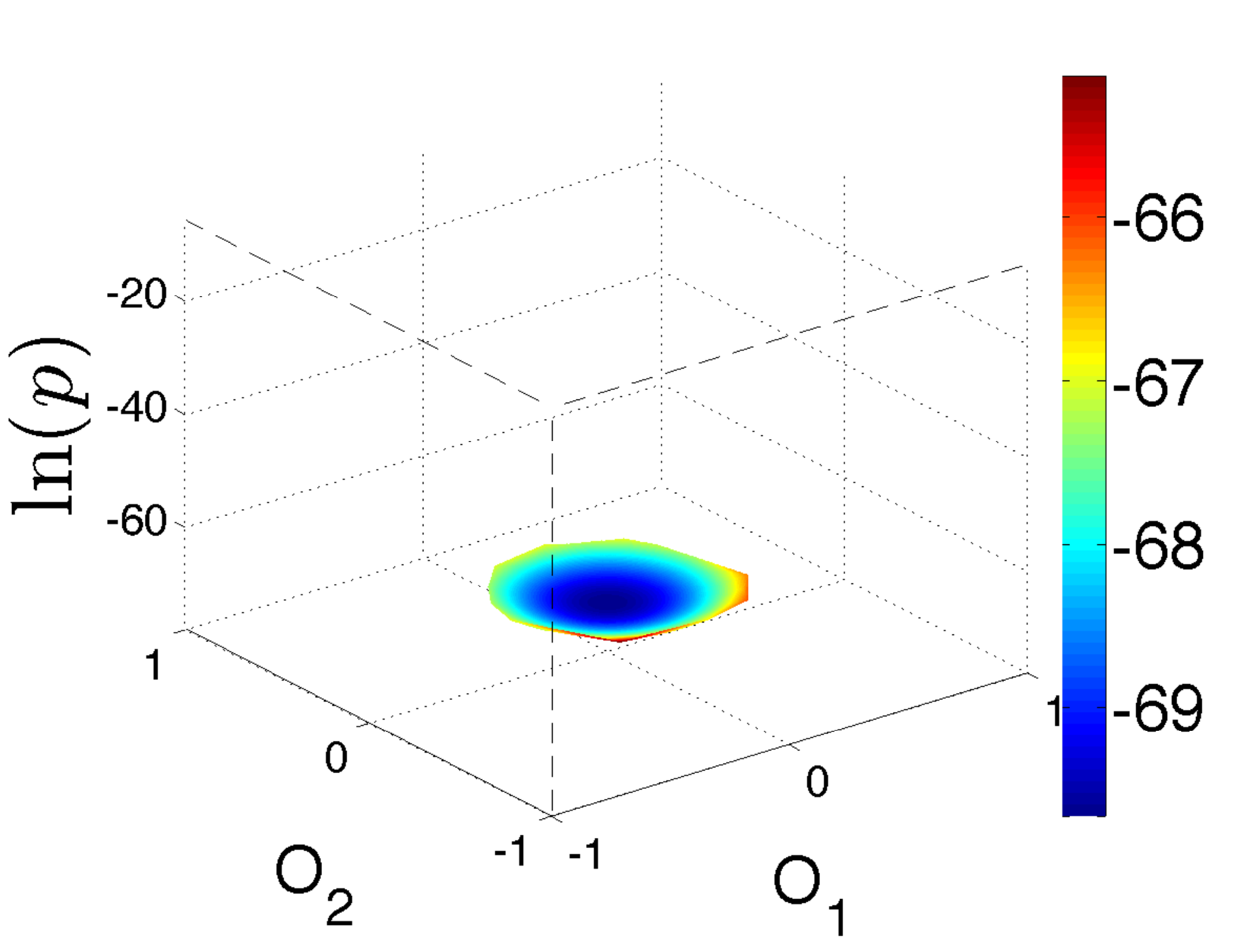}
\includegraphics[width=0.48\linewidth]{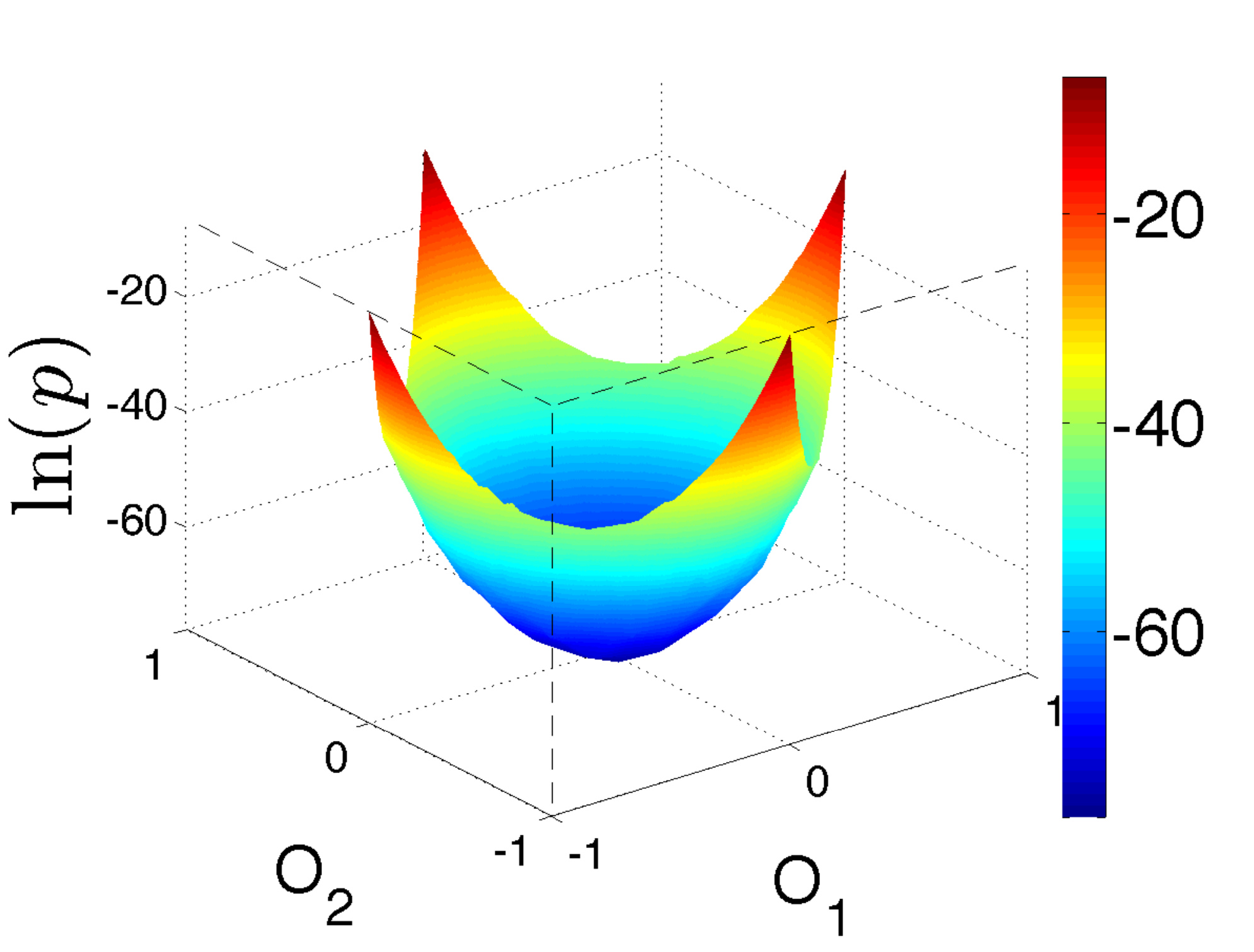}
\caption{Log probability of sampled configurations from VAN trained for a Hopfield model with $N = 100$ spins, and $P = 2$ orthogonal patterns. The sampled configurations are projected onto the two-dimensional space spanned by the two patterns. $X$ axis ($O_1$) and $Y$ axis ($O_2$) are the overlap (inner product, normalized to $[-1, 1]$) between each sampled configuration and the two patterns, respectively.
(a) $\beta = 0.3$, and the system is in the paramagnetic phase.
(b) $\beta = 1.5$, and the system is in the retrieval phase.
Note the different scales in the color bars.}
\label{fig:hop}
\end{figure}

Next, to demonstrate the ability of capturing multiple states at low temperature, we consider the Hopfield model~\cite{hopfield1982}, where $N$ spins are connected to each other. The couplings composed of $P$ random patterns,
$J_{i j} = \frac{1}{N} \sum_{\mu = 1}^P \xi_i^\mu \xi_j^\mu$,
with $\{\xi^\mu\} \in \{\pm 1\}^N$ denoting a random pattern.
At a low temperature with $P$ small, the system has a retrieval phase where all $P$ patterns are remembered by the system; hence there are $P$ pure states in the system~\cite{amit1985, amit1985a}.
The experiments are carried out on a Hopfield network with $N = 100$ spins and $P = 2$ orthogonal random patterns. At low temperature the energy (probability) landscape contains four modes, corresponding to two stored patterns and their mirrors (due to $\mathbb Z_2$ symmetry).
As opposed to models defined on lattices, there is no topology structure to apply convolution, so we use a simplest VAN with only one layer and $N (N - 1) / 2$ parameters.
We start training our network at $\beta = 0.3$ and slowly anneal the temperature to $\beta = 1.5$. At each temperature, we sample configurations from the trained VAN, and show their log probability in Fig.~\ref{fig:hop}.

The figure shows that at high temperature with $\beta = 0.3$, samplings are not correlated with the two stored patterns, and the system is in the paramagnetic state. The log probability landscape is quite flat, as the Gibbs measure is dominated by entropy.
When $\beta$ is increased to $1.5$, four peaks of probability emerge and dominate over other configurations. These four peaks touch coordinates $[1, 0]$, $[0, 1]$, $[-1, 0]$. and $[0, -1]$ in the $X$-$Y$ plane, which correspond exactly to the two patterns and their mirrors.
This is an evidence that our approach avoids collapsing into a single mode, and gives samplings capturing the features of the whole landscape, despite that those modes are separated by high barriers.

Compared with the landscape of Hopfield model in the retrieval phase which exhibits several local minima in the energy and probability landscape, models in the spin glass phase are considerably more complex~\cite{mezard1987}, because they have an infinite number of pure states, in the picture of replica symmetry breaking~\cite{parisi1980}.
Here we apply our method to the classic Sherrington-Kirkpatrick (SK) model~\cite{sherrington1975}, where $N$ spins are connected to each other by couplings $J_{i j}$ drawn from Gaussian distribution with variance $1/N$. So far the tensor network approaches do not apply to this model because of long range interactions and the disorder, which causes negative $Z$ issue~\cite{wang2014topological}. On the thermodynamic limit with $N \to \infty$ where the free energy concentrates to its mean value averaged over disorder, using for example replica method and cavity method, and replica symmetry breaking, i.e., the Parisi formula~\cite{parisi1980}.
On a single instance of SK model, the algorithm version of the cavity method, belief propagation, or Thouless--Anderson--Paler~\cite{thouless1977solution} equations apply as message passing algorithms. On large systems in the replica symmetry phase, the message passing algorithms converge and the obtained Bethe free energy is a good approximation, but in the replica symmetry breaking phase they fail to converge. Also notice that even in the replica symmetry phase, Bethe free energy is not an upper bound to the true free energy.

As a proof of concept, we use a small system size $N = 20$, so we can enumerate all $2^N$ configurations, compute the exact value of free energy, then evaluate the performance of our approach. Again, we use a simple VAN with only one layer.

\begin{figure}[tb]
\centering
\includegraphics[width=0.48\linewidth]{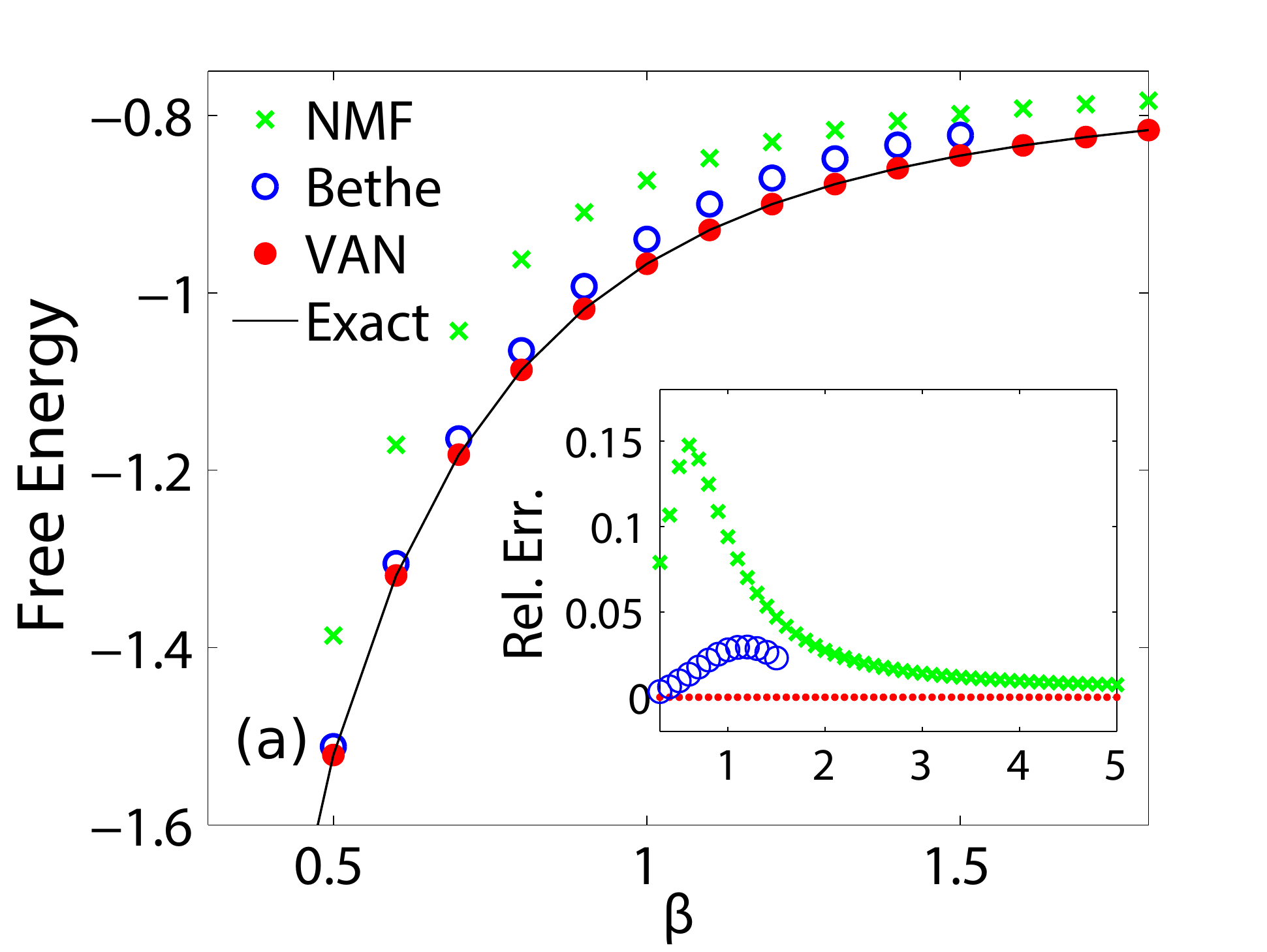}
\includegraphics[width=0.48\linewidth]{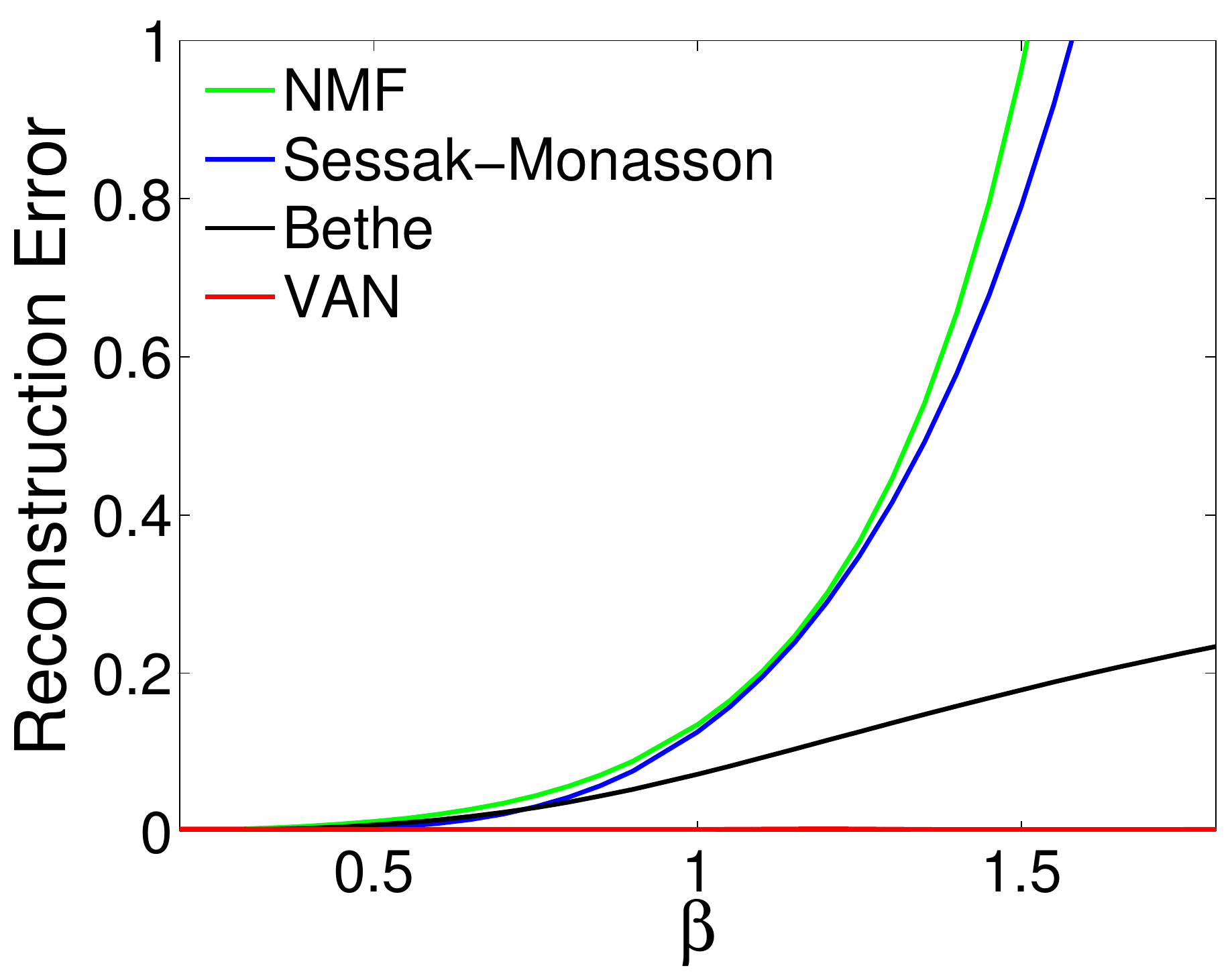}
\caption{(a) Free energy of SK model with $N = 20$ spins. The inset shows relative errors to exact values in a larger $\beta$ regime. Bethe converges only when $\beta \le 1.5$.
(b) The reconstruction error in the inverse Ising problem. The underlying model is an SK model with $N = 20$ spins. VAN uses a network with two layers (a hidden layer and an output layer).}
\label{fig:sk}
\end{figure}

In Fig.~\ref{fig:sk}(a) we show the free energy obtained from VAN, compared with NMF and Bethe approximations.
The free energy from VAN is much better than NMF and Bethe, and even indistinguishable to the exact value. This is quite remarkable considering that VAN adopts only $N (N - 1) / 2$ parameters, which is even smaller than that used in the belief propagation, $N (N - 1)$.
We also checked that our approach not only gives a good estimate on free energy, it also obtains accurate energy, entropy, magnetizations, and correlations.

The ability of solving \emph{ordinary} statistical mechanics problems also gives us the ability to solve \emph{inverse} statistical mechanics problems. A prototype problem is the \emph{inverse Ising problem}, which asks us to reconstruct the couplings of an Ising (spin glass) model, given the correlations~\cite{sm}.
It is well known that the Ising model is the maximum entropy model given the first and the second moments, so the couplings are uniquely determined by the correlations.
The problem has been studied for a long time especially in the field of statistical mechanics~\cite{nguyen2017inverse}, mainly using mean-field based methods.

The adaptation of our method for the inverse problem is straightforward by repeating the following two steps, until the correlations given by VAN are close enough to the given correlations of the underlying model: (1) train a VAN according to the Ising model with an existing $J_{i j}$ by minimizing the variational free energy; (2) compute correlations via direct sampling from the VAN, then update $J_{i j}$ to minimize the difference between the two sets of correlations.
We use our approach to reconstruct an SK model with $N = 20$ spins, and the given correlations are computed exactly by enumerating all $2^N$ configurations. The VAN uses two layers with $2000$ parameters.
The results are shown in Fig.~\ref{fig:sk}(b). Our method works much better than the popular mean-field methods of na\"ive mean-field~\cite{kappen1998, roudi2009a}, Sessak--Monasson small-correlation expansions~\cite{sessak2009}, and those based on a Bethe approximation~\cite{nguyen2012, ricci-tersenghi2012}, especially in the glassy phase with $\beta > 1$.

\paragraph{Outlooks.}

In the present Letter, we have focused on binary spins. However, it is straightforward to generalize the approach to Potts models and models with continuous variables. We also notice that, for continuous variables and with a regular structure, a flow-based model together with a renormalization group has been proposed for the variational free energy minimization problem~\cite{classic-mera}.
For systems defined on a $2$D lattice we have shown how to adopt convolutions for respecting the $2$D structure of the underlying factor graph~\cite{pixelcnn}. This strategy can be extended straightforwardly to systems on $3$D lattices using $3$D convolutions, and to graphical models on an arbitrary factor graph using e.g. graph convolution networks~\cite{bronstein2017geometric} with proper filters.

We anticipate that our method will find immediate applications in a broad range of disciplines. For example, it can be applied directly to statistical inference problems, where the Boltzmann distribution in statistical mechanics becomes the posterior distribution of Bayesian inference~\cite{zdeborova2016statistical}. Another example of application would be the combinatorial optimizations and constraint satisfaction problems, in which finding the optimal configurations and solutions correspond to finding ground states of spin glasses, and counting the number of solutions corresponds to computing entropy at zero temperature.

So far our approach is rather a proof of concept of a promising variational framework on statistical physics problems.
Building on the current work, an interesting direction for future work would be even more deeply incorporating successful physics and machine learning concepts (such as a renormalization group and dilated convolution) into the network architecture design, e.g., the \textsc{WaveNet}~\cite{wavenet}. This would allow us to scale to a much larger problem size, or even to the thermodynamic limit.

The main limitation of our method is that the variational free energy calculation relies on sampling of the model; hence it is slower than canonical variational mean-field message passing algorithms which compute variational free energy directly using model parameters. We also notice that the sampling process can be sped up by caching intermediate activations in the sampling procedure as explored in Refs.~\cite{paine2016fast,ramachandran2017fast}. Or, one may use alternative model such as inverse autoregressive flow~\cite{kingma2016}, which supports parallel sampling.

A pytorch implementation of our model and algorithms is avaliable at Ref.~\cite{github}.

\begin{acknowledgments}
We thank Zhiyuan Xie and Haijun Zhou for discussions.
L.W. is supported by the Ministry of Science and Technology of China under the Grant No. 2016YFA0300603, the National Natural Science Foundation of China under the Grant No. 11774398, and the Strategic Priority Research Program of Chinese Academy of Sciences Grant No.~XDB28000000.
P.Z. is supported by Key Research Program of Frontier Sciences, CAS, Grant No. QYZDB-SSW-SYS032 and Project 11747601 of National Natural Science Foundation of China.
\end{acknowledgments}


%

\newpage
\phantom{blahblah}
\newpage

\appendix

\section{Autoregressive networks}

The key feature of the autoregressive model is the ability of computing normalized probability of a spin configuration $\s$ (which is the input to the autoregressive network). This ability comes from the design of the joint probability distribution
\begin{equation}
q_\theta(\s) = \prod_{i = 1}^N q_\theta(s_i \mid s_1, \ldots, s_{i - 1}).
\label{eq:autoregressive_}
\end{equation}
As a simple example, the joint probability of $4$ variables $\{s_1, s_2, s_3, s_4\}$ can be evaluated using Bayes rule as
\begin{align}
&\phantom{{}={}} p(s_1, s_2, s_3, s_4) \nonumber \\
&= p(s_4 \mid s_1, s_2, s_3) p(s_1, s_2, s_3) \nonumber \\
&= p(s_4 \mid s_1, s_2, s_3) p(s_3 \mid s_1, s_2) p(s_1, s_2) \nonumber \\
&= p(s_4 \mid s_1, s_2, s_3) p(s_3 \mid s_1, s_2) p(s_2 \mid s_1) p(s_1).
\end{align}
The autoregressive network essentially approximates all the conditional probabilities using neural networks with a polynomial number of parameters. Notice that exactly storing those conditional probabilities requires an exponential number of parameters in the worst case.

For the spin variables, we choose sigmoid function (ranging in $(0, 1)$) as an activation function, and the output of the network is given by
\begin{equation}
\hat s_i = \sigma \left( \sum_{j < i} W_{i j} s_j \right).
\end{equation}
To compute the likelihood of a given configuration $\s$, one pass the configuration to the model to compute $\hat{\s}$, based on which one has the log likelihood given by
\begin{equation}
\ln q_\theta(\s) = \sum_{i = 1}^N \ln \left[ \hat s_i \delta_{s_i, +1} + (1 - \hat s_i) \delta_{s_i, -1} \right].
\label{eq:q_}
\end{equation}
Evidently, the exact expression of $\ln q_\theta(\s)$ enables us to compute the variational free energy from unbiased samples of the model, that is
\begin{equation}
F_q = \frac{1}{\beta} \sum_\s q_\theta(\s) \left[ \beta E(\s) + \ln q_\theta(\s) \right],
\label{eq:vf_}
\end{equation}
where $E(\s) = \sum_{(i j)} J_{i j} s_i s_j$ is defined by the Ising model, and the $q_\theta(\s)$ is from \Eq{eq:q_}.

Another essential technique followed from the above expression for computing the variational free energy is how to obtain many unbiased samples drawn from the model. Fortunately, this is easy thanks to the design of the autoregressive networks. Since we have stored all approximated joint conditional probabilities, sampling of the autoregressive model directly follows the factorization of the conditional probabilities, following the predetermined order starting from the first to the last one.
To illustrate the sampling process, let us again use the simple example of
\begin{equation}
q_\theta(\s) = p(s_4 \mid s_1, s_2, s_3) p(s_3 \mid s_1, s_2) p(s_2 \mid s_1) p(s_1).
\end{equation}
To sample a configuration $\s = \{s_1, s_2, s_3, s_4\}$, we first toss a coin to determine $s_1$ using marginal probability $p(s_1)$, then toss a coin again to determine $s_2$ using $p(s_2 \mid s_1)$. The assignment of $s_3$ and $s_4$ can be determined in turn.

\section{Gradient estimator and variance reduction}

Given the samples drawn from the model, and the variational free energy $f(\theta)$ computed using the samples, one thing we need to be careful is that one cannot compute directly the gradient of $f$ with respect to model parameters $\theta$. Instead, one must derive a proper gradient estimator, which is written as
\begin{align}
\beta \nabla_\theta F_q &= \nabla_\theta \sum_\s \left[ q_\theta(\s) \cdot \left( \beta E(\s) + \ln q_\theta(\s) \right) \right] \nonumber \\
&= \sum_\s \left[ \nabla_\theta q_\theta(\s) \cdot \left( \beta E(\s) + \ln q_\theta(\s) \right) + q_\theta(\s) \nabla_\theta \ln q_\theta(\s) \right] \nonumber \\
&= \mathbb{E}_{\s \sim q_\theta(\s)} \left[ \nabla_\theta \ln q_\theta(\s) \cdot \underbrace{\left( \beta E(\s) + \ln q_\theta(\s) \right)}_{R(\s)} \right],
\end{align}
which is \Eq{eq:gradient} in the main texts.
The contribution of each sample $\s \sim q_\theta(\s)$ to the gradient is $\nabla_\theta \ln q_\theta(\s)$ weighted by the reward signal $R(\s)$. This ensures that when $R(\s)$ is large, the optimizer will try to reduce the probability of generating such configuration, hence reduce the variational free energy. Learning the probability distribution using this score function gradient estimator is also known as the REINFORCE algorithm~\cite{williams1992simple} in reinforcement learning literatures~\cite{sutton1998reinforcement}.

We also notice that in deriving the last equation we have used
\begin{equation}
\mathbb{E}_{\s \sim q_\theta(\s)} \left[ \nabla_\theta \ln q_\theta(\s) \right] = \nabla_\theta \sum_\s q_\theta(\s) = \nabla_\theta 1 = 0.
\end{equation}
For the same reason, one can subtract any $\s$-independent constant in the last equation without affecting the expectation, that is
\begin{align}
\nabla_\theta F_q
&= \frac{1}{\beta}\mathbb{E}_{\s \sim q_\theta(\s)} \left[ \nabla_\theta \ln q_\theta(\s) \cdot (R(\s) - b) \right].
\end{align}
The baseline $b$ is useful to reduce the variance of the gradient, and is known as \emph{variance reduction} in the context of the reinforcement learning literature~\cite{dlbook}.
In this work we consider only a simplest strategy~\cite{mnih2014neural} by setting the baseline to
\begin{equation}
b = \mathbb{E}_{\s \sim q_\theta(\s)} R (\s).
\end{equation}
$b = \mathbb{E}_{\s \sim q_\theta(\s)} R (\s)$ is the estimate of log partition function. It is computed at each iteration. There is a simple understanding of this choice of baseline: the actual loss function can be written as
\begin{equation}
\mathcal{L} = \mathbb{E}_{\s \sim q_\theta(\s)} \left[ \beta E(\s) + \ln q_\theta(\s) - \ln \tilde Z \right] \approx D_\text{KL}(q_\theta \,\|\, p_{\mathrm{Boltzmann}}),
\end{equation}
where $\ln \tilde Z = b$ is the estimated log partition function. Optimizing $\mathcal{L}$ is equivalent to optimizing the variational free energy, but with the advantage of reduced variance which was induced by the gap between magnitudes of $E(\s)$ and $\ln q_\theta(\s)$.

\section{Zero variance condition and exact free energy}

The variational free energy is an estimator over the variational distribution,
\begin{equation}
F_q = \mathbb{E}_{\s \sim q_\theta(\s)} \left[ E(\s) + \frac{1}{\beta} \ln q_\theta(\s) \right].
\label{eq:mean}
\end{equation}
If the exact distribution is achieved, i.e. $q_\theta(\s) = p_\text{Boltzmann}(\s)$, we have
\begin{equation}
E(\s) + \frac{1}{\beta} q_\theta(\s) = -\frac{1}{\beta} \log(Z),
\end{equation}
which means that the quantity $E(s) + \frac{1}{\beta} \ln q_\theta(\s)$ has \emph{zero variance}.

\begin{figure}[tb]
\centering
\includegraphics[width=0.8\linewidth]{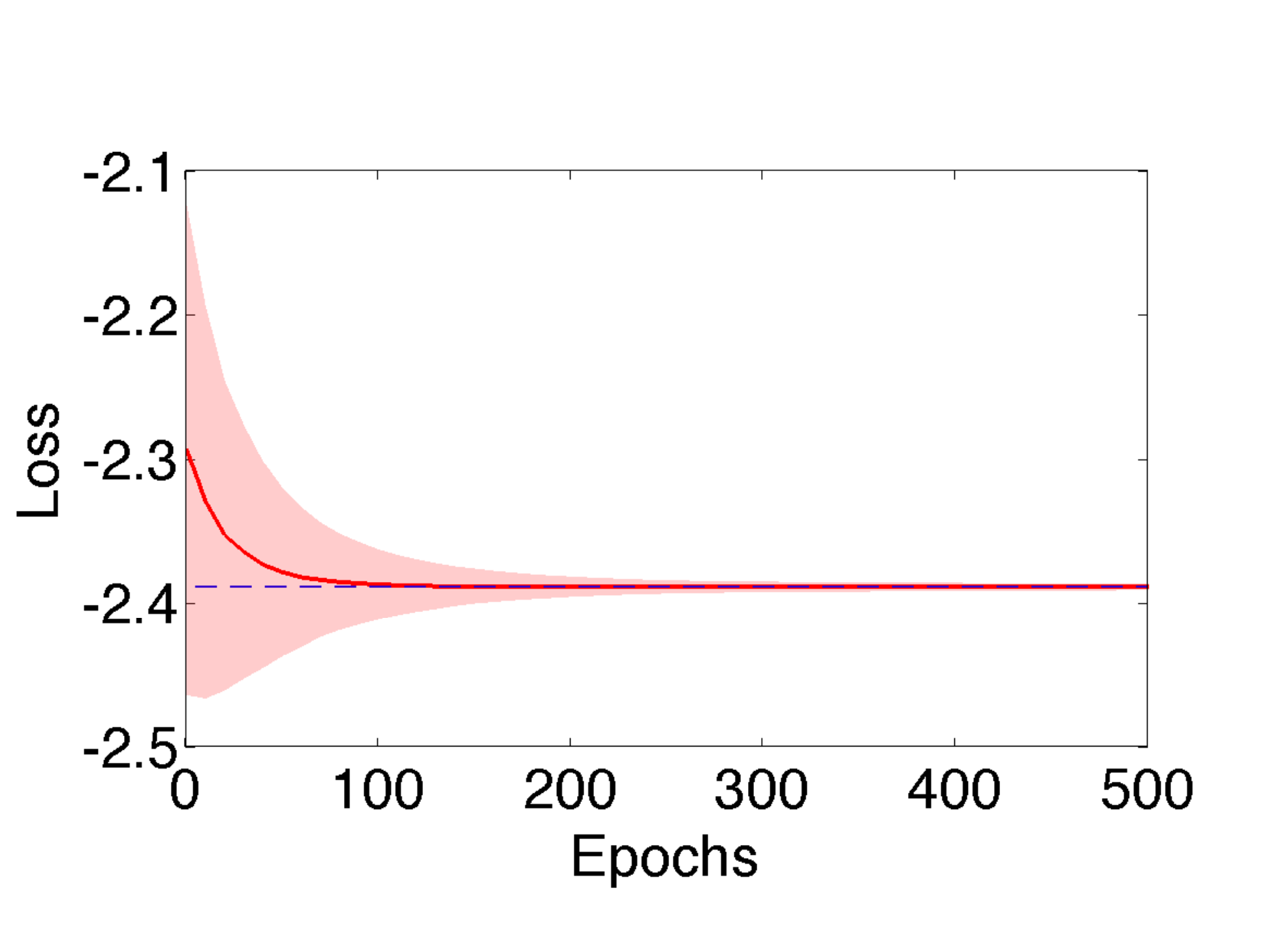}
\caption{Evolution of mean and variance of the loss function during the training of VAN on an SK model with $N = 20$ spins, $\beta = 0.3$. The light red area denotes variance, the red line denotes mean, and the blue dashed line denotes the exact free energy value. The VAN uses $2$ layers and totally $8200$ trainable parameters.}
\label{fig:var}
\end{figure}

On the other hand, if the variance is zero, then the distribution $q_\theta$ must be a Boltzmann distribution. To prove this, we make use that zero variance implies the quantity in the square bracket to be a constant,
\begin{equation}
E(\s) + \frac{1}{\beta} \ln q_\theta(\s) = C.
\end{equation}
Solving the equation gives \begin{equation}
q_\theta(\s) = \e^{\beta C - \beta E(\s)}.
\end{equation}
However, we note that the $\e^{\beta C}$ does not necessarily be equal to $1/Z$, the normalization of the original Boltzmann distribution, due to \emph{mode collapse} where not all the modes (pure states) of the original Boltzmann distribution are captured by the network.

Nevertheless, if one can ensure that mode collapse never happens, then low variance indeed indicates a good estimate to the true free energy. In this work we propose to use temperature annealing to avoid mode collapse, and our results in Fig.~\ref{fig:hop} in the main texts give an evidence that mode collapse does not happen.
Therefore, we can use the variance to practically indicate the closeness between $q_\theta$ and the exact distribution without knowing the latter.
As an example to illustrate this, in Fig.~\ref{fig:var} we plot evolution of $F_q$ from \Eq{eq:mean} and the variance during training of a VAN on an SK model with $N = 20$ spins. The figure shows that when the variance (light red area) decreases during training, the variational free energy (red line) converges to the true free energy (blue dashed line).

\section{Inverse Ising problem and conventional mean-field methods}

It is well known that the Ising (spin glass) model
\begin{equation}
p(\s) = \frac{1}{Z} \e^{\sum_{i j} J_{i j} s_i s_j + \sum_i h_i s_i}
\end{equation}
is the maximum entropy model when the first and the second moments of the distribution $p(\s)$ are constrained. The inverse Ising problem asks to reconstruct external fields $\{h_i\}$ and couplings $\{J_{i j}\}$ of the underlying model when magnetizations $\{m_i\}$ and correlations $\{C_{i j}\}$ are given, where
\begin{align}
m_i &= \sum_{\s} p(\s) s_i, \\
C_{i j} &= \sum_{\s} p(\s) s_i s_j.
\end{align}
The maximum likelihood inference gives a simple condition that the magnetizations $m(\mathbf{h}, \mathbf{J})$ and the correlations $C(\mathbf{h}, \mathbf{J})$ from the trained network should match the given ones. If they do not match, the difference between the two quantities provide gradient for learning external fields and couplings.

The main difficulty of reconstruction is that exact magnetizations and correlations from the trained network (i.e. with learned external fields and couplings) are computationally intractable. Various mean-field methods have been proposed for estimating them. In our method, we estimate magnetizations and correlations using configurations sampled from the trained autoregressive network, which provides efficient direct sampling.

In this paper, we consider models with no external field, thus the task is to reconstruct couplings from correlations. Magnetizations are exactly zero, due to the $\mathbb{Z}_2$ symmetry. To avoid the influence of measurement noise in correlation data, we test our method on small systems and compute exact correlations by enumerating all the configurations. We compare the performance of our method with several well-known mean-field methods, including na\"ive mean-field method (NMF), Sessak--Monasson small correlation expansion method and Bethe approximation. In NMF, the correlations are computed using na\"ive mean-field approximation and linear response relation~\cite{kappen1998boltzmann}, and coupling are given by
\begin{equation}
J^\text{NMF}_{i j} = \delta_{i j} - (C^{-1})_{i j}.
\end{equation}
In Sessak--Monasson small correlation expansion, a perturbation expansion of entropy in terms of the connected correlation is carried out, and the reconstructed couplings are given by
\begin{equation}
J^\text{SM}_{i j} = -(C^{-1})_{i j} + J_{i j}^\text{IP} - \frac{C_{i j}}{1 - C_{i j}^2},
\end{equation}
where
\begin{equation}
J^\text{IP}_{i j} = \frac{1}{4} \ln \left[ \frac{(1 + C_{i j})^2}{(1 - C_{i j})^2} \right]
\end{equation}
is known as the independent-pair approximation.
Bethe approximation~\cite{nguyen2012, ricci-tersenghi2012} is rather simple,
\begin{equation}
J^\text{Bethe} = \frac{1}{2} \mathrm{arcsinh} \left[ 2 (C^{-1})_{i j} \right].
\end{equation}
The key ingredient in deriving the last formula is a careful computation of correlations given by Bethe approximation. One method is the sophisticated susceptibility propagation algorithm~\cite{mezard2009a} which computes the connected correlations by applying the linear response relation to the belief propagation~\cite{mezard2009a}.
We refer to~\cite{nguyen2017inverse} for an overview of these mean-field methods.

After all, the performance of reconstruction is characterized by the reconstruction error between the inferred couplings $J_\text{infer}$ and true couplings $J_\text{true}$, defined as
\begin{equation}
\Delta_J = \frac{1}{N} \sqrt{\sum_{i j} \left( J^\text{infer}_{i j} - J^\text{true}_{i j} \right)^2}.
\end{equation}

\section{More results on Sherrington--Kirkpatrick model, Hopfield model, and the inverse SK model}

\begin{figure*}[tb]
\centering
\includegraphics[width=0.32\linewidth]{fig/hop3.pdf}
\includegraphics[width=0.32\linewidth]{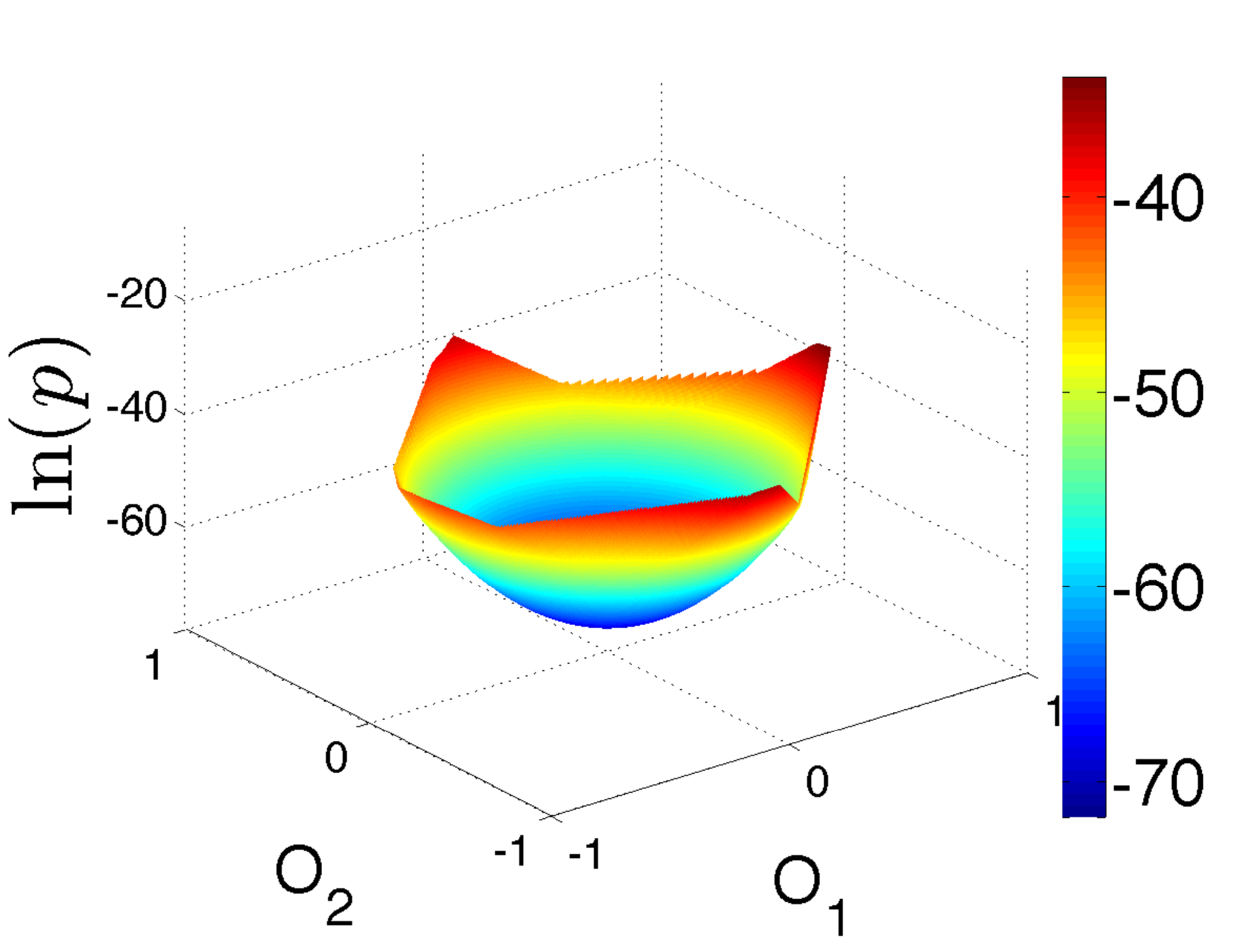}
\includegraphics[width=0.32\linewidth]{fig/hop1.pdf}
\includegraphics[width=0.32\linewidth]{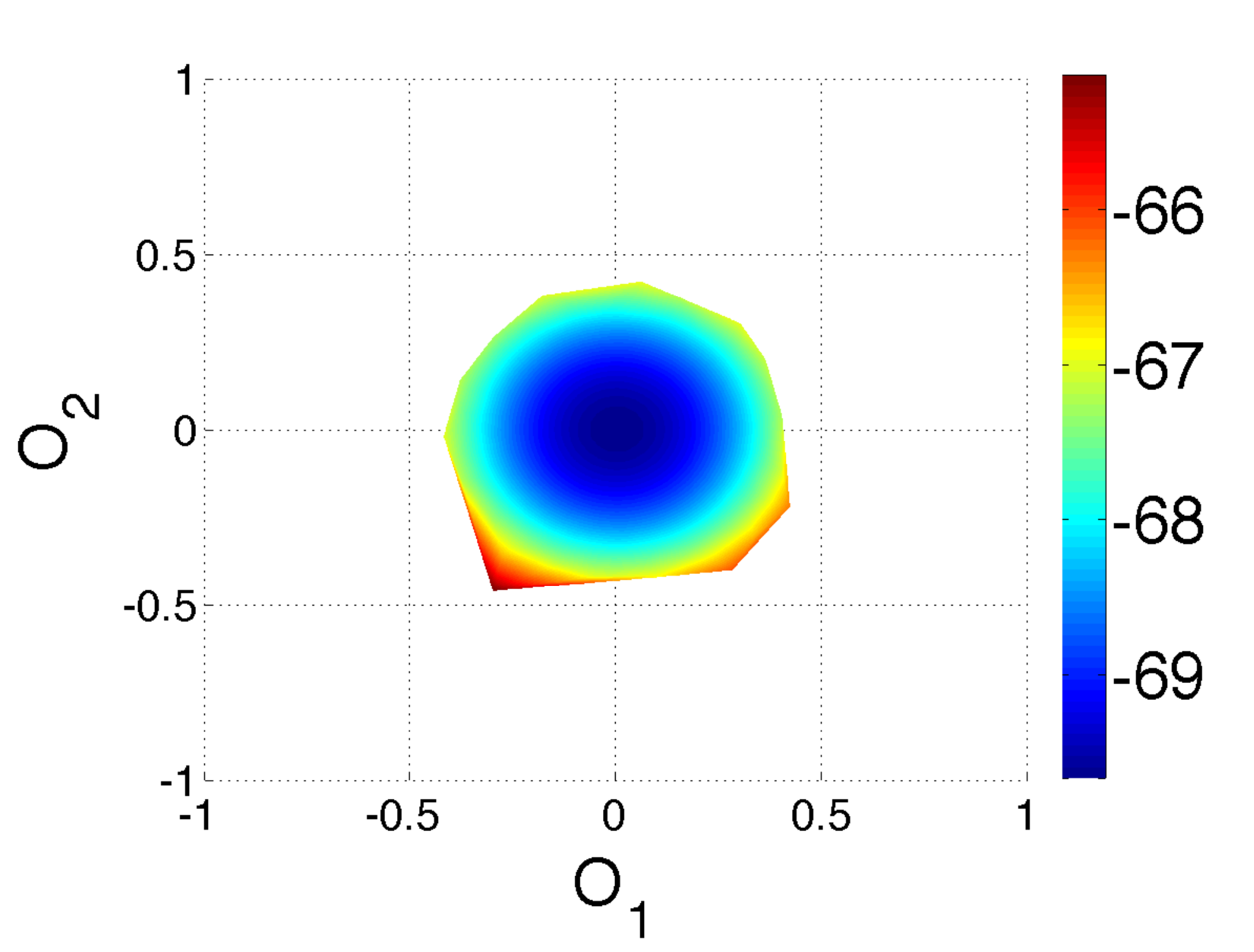}
\includegraphics[width=0.32\linewidth]{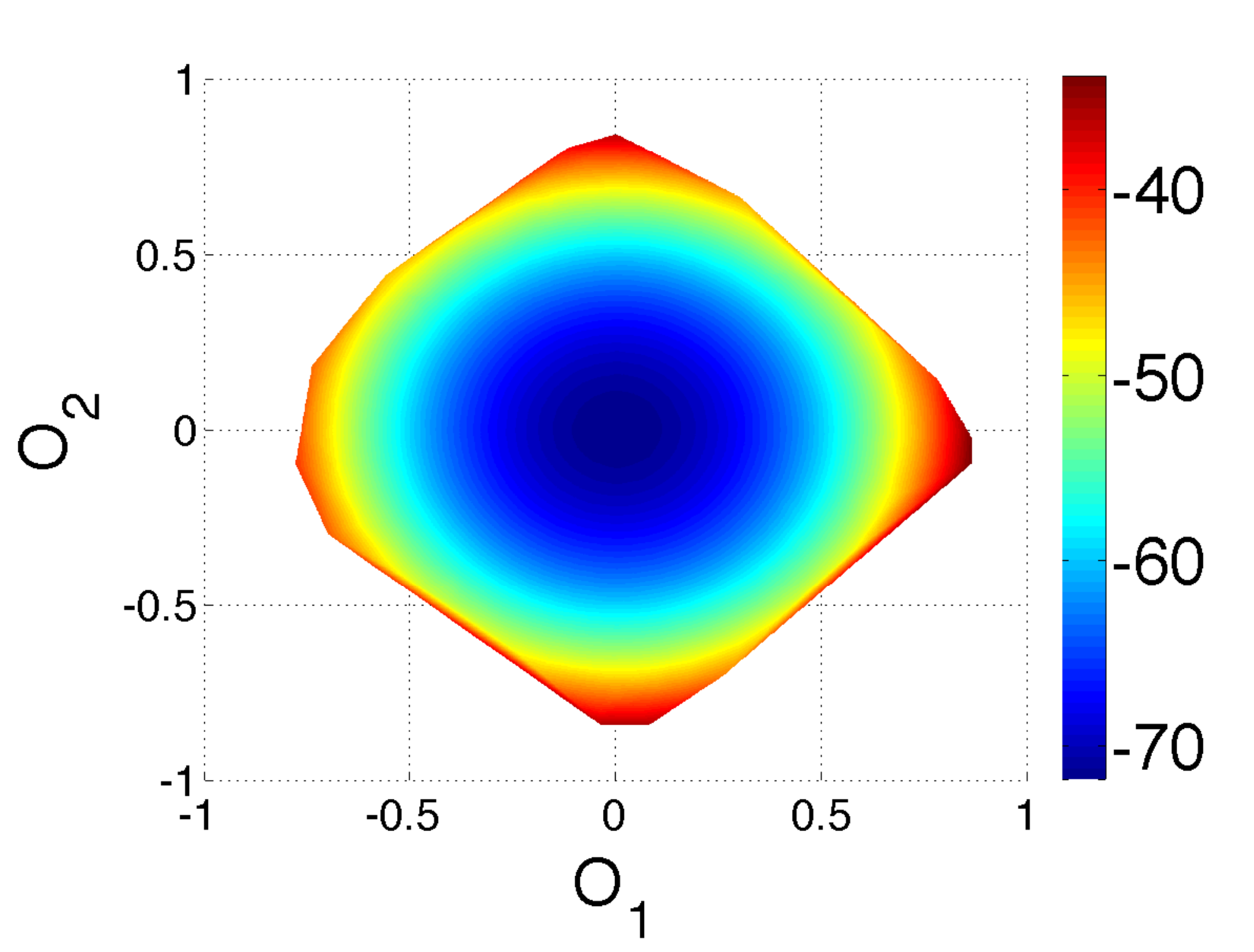}
\includegraphics[width=0.32\linewidth]{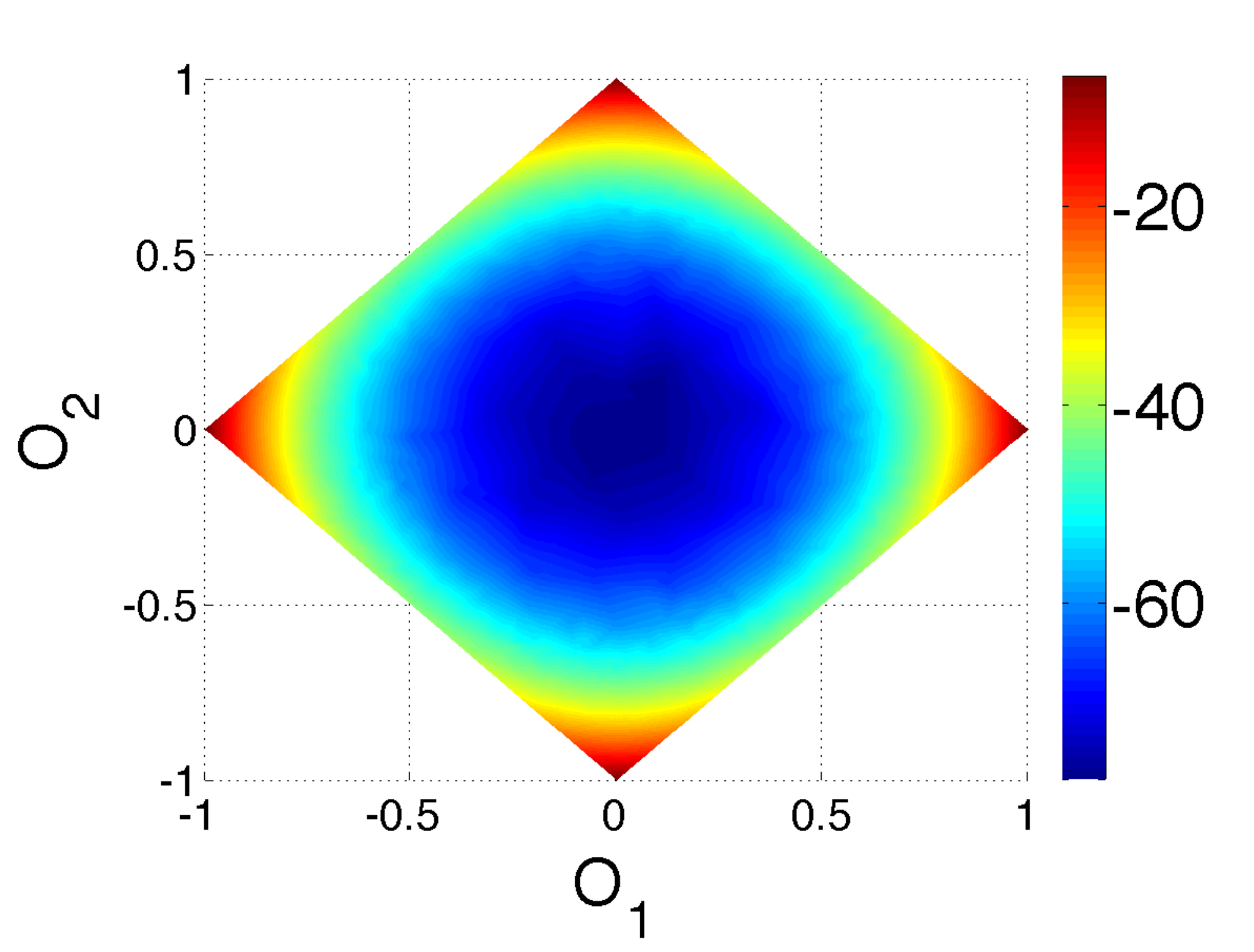}
\caption{Log probability of sampled configurations of VAN learned for a Hopfield model with $N = 100$ spins, and $P = 2$ orthogonal patterns, on the two-dimensional spaces spanned by two patterns. In the figures, $X$-axis and $Y$-axis represent inner product (overlap) between each sampled configuration and the first and the second stored patterns respectively. Our network uses single layer and only $N (N - 1) / 2$ parameters. The top figures are $3$D view of the meshed log probability surface, while the bottom figures are the $2$D view from top. From left to right, $\beta$ values are $0.3$, $1.0$, and $1.5$ respectively. }
\label{fig:hop_more}
\end{figure*}

In Fig.~\ref{fig:hop_more} we draw a more detailed process of annealing, depicted by the landscape of log probability of samples (which is equivalent to the energy landscape up to a constant), in both $3$D and $2$D views. From the figures we can see that at a high temperature, with $\beta$ small, samples are of uniform measure, roughly around $-N \ln(2)$. This is because the system is in the paramagnetic phase with a single paramagnetic mode, thus there should be no mode collapse. As $beta$ increases to $1.0$, which is the spin-glass transition point at the thermodynamic limit, from the middle panels we can recognize $4$ little peaks, corresponding to two stored patterns and their mirrors, begin to emerge a little. In this situation the modes are rather weak (at the transition point), and they are quite easy to capture by VAN. Once those little peaks are captured, they grow as the temperature decreases gradually, and finally arrive at sharp peaks at a low temperature, as shown in the right panels in the figure. Although this is not a proof that annealing works sufficiently to avoid mode collapse, We think the phenomenon displayed at $\beta = 1.0$ explains why and how annealing works in alleviating mode collapse.

\begin{figure}[tb]
\centering
\includegraphics[width=0.6\linewidth]{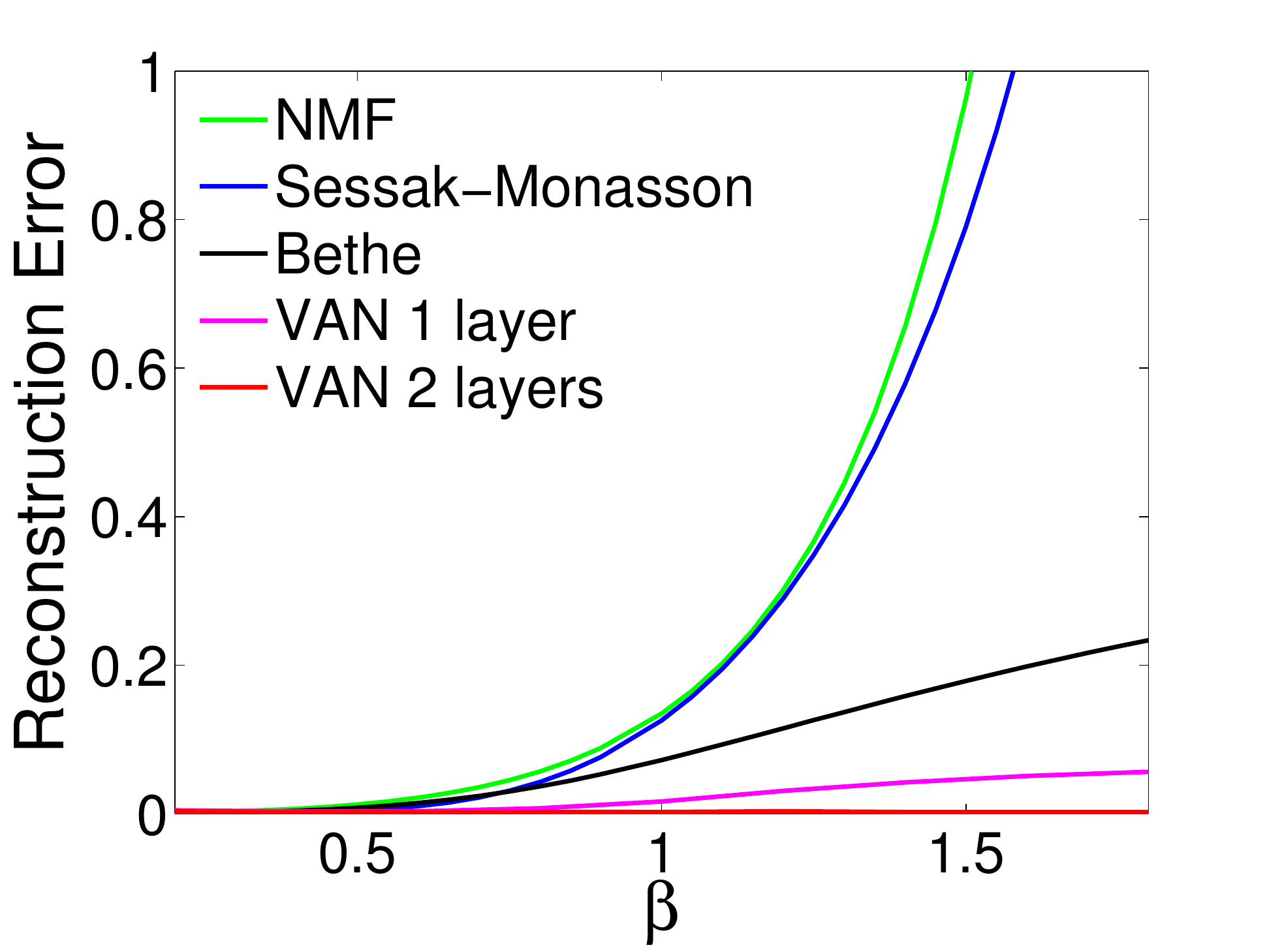}
\caption{The reconstruction error in the inverse Ising problem compared with mean-field methods. The underlying model is an SK model with $N = 20$ spins. VAN $1$ layer uses an autoregressive network with one layer (there is no hidden layer), and $190$ parameters. VAN $2$ layers uses an autoregressive network with a hidden layer and an output layer, with totally $2000$ parameters.}
\label{fig:invsk}
\end{figure}

In Fig.~\ref{fig:invsk} we added VAN results with only $1$ layer and very few parameters.

\section{Calculation of heat capacity and critical temperature for Ising model}

\begin{figure}[tb]
\centering
\includegraphics[width=0.6\linewidth]{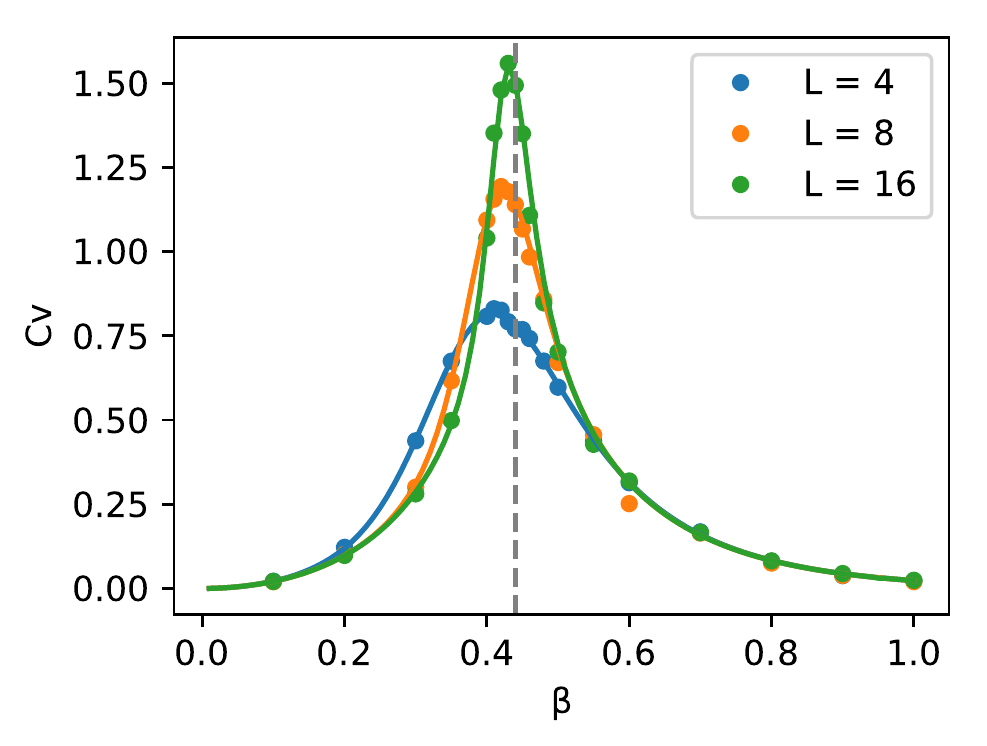}
\caption{
Heat capacity per site of ferromagnetic Ising model on square lattices with periodic boundary condition. Curves are exact values, markers are obtained using VAN, and the vertical dashed line indicates the critical point $\beta_c = \frac{1}{2} \ln \left( 1 + \sqrt{2} \right)$ for $L \to \infty$.
}
\label{fig:fm-sqr-cv}
\end{figure}

The ferromagnetic Ising model on infinitely large $2$D square lattice shows phase transition at the critical temperature $\beta_c$, where the heat capacity $C_v$ goes to infinity. The heat capacity can be calculated by
\begin{equation}
C_v = \beta^2 \mathrm{Var}[E] = \beta^2 \left( \langle E^2 \rangle - \langle E \rangle^2 \right),
\end{equation}
where the expectation $\langle \cdot \rangle$ is computed through samples from the variational distribution. For every finite and fixed $L$, we plot $C_v$ against $\beta$, find the peak position $\beta_c(L)$, and calculate the critical exponent by fitting the power function. The heat capacity for $L = 4, 8, 16$ given by VAN are shown in Fig.~\ref{fig:fm-sqr-cv}.

In principle we can obtain the critical temperature and critical exponents for the infinitely large system by extrapolating $L \to \infty$. However, running VAN with larger $L$ becomes computationally expensive, and it remains an open problem how to systematically change the network size to trade off between speed and precision.

\section{Details on network structure and training for Ising models}

For $2$D Ising model, we set the lattice size to be $16 \times 16$, and specify the network's depth (the number of layers) and width (the number of channels in a layer). We test convolution layers and densely connected layers respectively. For convolution layers, we also specify the kernel radius ($\text{kernel radius} \times 2 + 1 = \text{edge length of the kernel}$). To cover a lattice with edge length $L$, depth and kernel radius should satisfy
\begin{equation}
\text{depth} \times \text{kernel radius} + 1 \ge L.
\label{eq:cover}
\end{equation}

We test a ``shallow'' network with depth = $3$, and a ``deep'' network with depth = $6$ and residue blocks~\cite{resnet}. The result shown in the main text is chosen according to a lower free energy between them. In practice, we find that the ``shallow'' one gives lower free energy for high temperature, otherwise the ``deep'' one works better. For convolution layers, we set width = $64$, because it is the elbow point when we plot the relative error of the free energy versus the number of parameters. The values of depth, width, kernel radius and number of parameters are summarized in Table~\ref{tab:hyperparam-ising}. These network sizes ensure the number of parameters in different networks are within the same magnitude.

\begin{table}[tb]
\addtolength{\tabcolsep}{12pt}
\centering
\begin{tabular}{ccc}
\toprule
& Shallow & Deep \\
\midrule
Conv & $3 / 64 / 6 / 714,113$ & $6 / 64 / 3 / 826,369$ \\
Dense & $3 / 4 / \text{-} / 1,577,216$ & $6 / 2 / \text{-} / 2,366,720$ \\
\bottomrule
\end{tabular}
\caption{Depth / width / kernel radius / number of parameters in employed networks.}
\label{tab:hyperparam-ising}
\end{table}

To implement $\mathbb Z_2$ symmetry, we create a mixture model of the network and itself with input inversed. The probability of the configuration $\s$ is $q_{\mathbb {Z}_2}(\s) = \frac{1}{2} \left( q_\theta(\s) + q_\theta(-\s) \right)$~\cite{moore2016symmetrized, classic-mera}, where $q_\theta(\s)$ is the probability given by the network. In sampling, we first generate a batch of samples from the network, then randomly inverse them by probability $1/2$.

We use Adam optimizer~\cite{adam} to minimize the variational free energy. To avoid mode collapse, we start training at infinite temperature ($\beta = 0$), and slowly increase $\beta$ until the desired temperature is reached. Moreover, we clip the norm of the gradient to increase the stability of training. We train $10,000$ steps to ensure the optimization converges, calculate the moving average of the free energy in $100$ steps, and report the lowest one.

The result of the variational free energy is insensitive to many of the hyperparameters. Numerical experiments show that if we change those hyperparameters in a range, the result will not change significantly. The hyperparameters we use in the reported results are shown in Table~\ref{tab:hyperparam-range-ising}.

\begin{table}[tb]
\centering
\begin{tabular}{ccc}
\toprule
Hyperparameter & Reported & Range \\
\midrule
Batch size & $1000$ & $100 \sim 10,000$ \\
Learning rate & $10^{-3}$ & $10^{-4} \sim 10^{-3}$ \\
Adam $\beta_1$ & $0.9$ & $0.5 \sim 0.99$ \\
Adam $\beta_2$ & $0.99$ & $0.9 \sim 0.99$ \\
Temperature annealing rate & $0.998$ & $0.99 \sim 0999$ \\
Gradient clipping norm & $1$ & $1 \sim 10$ \\
Data type & float32 & float32, float64 \\
\bottomrule
\end{tabular}
\caption{Hyperparameters used in reported results and their insensitive ranges.}
\label{tab:hyperparam-range-ising}
\end{table}

The typical one-step training time for $16 \times 16$ Ising model is $1.8$ sec. for the ``shallow'' network, and $5.4$ sec. for the ``deep'' network, on a single NVIDIA Titan V GPU.
Such large networks produce the lowest relative errors of free energy and physical observables, including energy and heat capacity, that are capable under our computation resources. If we merely want to outperform previous mean-field methods, we can use a much smaller network, and achieve faster training speed.

\begin{table}[tb]
\centering
\begin{tabular}{cc}
\toprule
Hyperparameters & Time of $1$ step \\
\midrule
$2$D Ising model, $L^2 = 256$ spins, & \\
$1$ layer, $1$ channel, $3 \times 3$ receptive field, & $0.076$ sec. \\
$10$ parameters, batch size $1000$ & \\
\midrule
$2$D Ising model, $L^2 = 256$ spins, & \\
$3$ layers, $64$ channels, $13 \times 13$ receptive field, & $1.8$ sec. \\
$714,113$ parameters, batch size $1000$ & \\
\midrule
$2$D Ising model, $L^2 = 256$ spins, & \\
$6$ layers, $64$ channels, $7 \times 7$ receptive field, & $5.4$ sec. \\
$826,369$ parameters, batch size $1000$ & \\
\midrule
$2$D Ising model, $L^2 = 256$ spins, & \\
$3$ layers, $4$ channels, dense connection, & $0.27$ sec. \\
$1,577,216$ parameters, batch size $1000$ & \\
\midrule
$2$D Ising model, $L^2 = 256$ spins, & \\
$6$ layers, $2$ channels, dense connection, & $0.50$ sec. \\
$2,366,720$ parameters, batch size $1000$ & \\
\bottomrule
\end{tabular}
\caption{Hyper parameters and one-step (epoch) training time of VAN on the $2$D Ising model, on a single NVIDIA Titan V GPU.}
\label{tab:time-2d}
\end{table}

\section{Details on network structure and training for SK model}

For the SK model, we have used batch size $10,000$, learning rate $0.001$, an input layer with $20$ neurons and an output layer with $20$ neurons.
For the inverse SK model, we have tried two kinds of VAN. The first one uses an input layer with $20$ neurons and an output layer with $20$ neurons. The other one uses additionally a hidden layer with $100$ neurons. We have used batch size $10,000$, learning rate for the inner loop (for learning VAN weights) $0.001$ and for the outer loop (for learning couplings) $0.01$.

The one-step training times for $2$D Ising model and SK model with several hyperparameters are shown in Table~\ref{tab:time-2d} and \ref{tab:time-sk}.

\begin{table}[tb]
\centering
\begin{tabular}{cc}
\toprule
Hyperparameters & Time of $1$ step \\
\midrule
SK model, $N = 20$ spins, & \multirow{2}{*}{$0.0082$ sec.} \\
$1$ layer, $190$ parameters, batch size $10,000$ & \\
\midrule
SK model, $N = 20$ spins, & \\
$2$ layers, $100$ hidden neurons, & $0.012$ sec. \\
$2000$ parameters, batch size $10,000$ & \\
\midrule
SK model, $N = 100$ spins, & \multirow{2}{*}{$0.030$ sec.} \\
$1$ layer, $4950$ parameters, batch size $10,000$ & \\
\midrule
SK model, $N = 100$ spins, & \\
$2$ layers, $500$ hidden neurons, & $0.065$ sec. \\
$25,000$ parameters, batch size $10,000$ & \\
\midrule
SK model, $N = 100$ spins, & \multirow{2}{*}{$0.10$ sec.} \\
$1$ layer, $4950$ parameters, batch size $100,000$ \\
\bottomrule
\end{tabular}
\caption{Hyper parameters and one-step (epoch) training time of VAN on the SK model, on a single NVIDIA Titan V GPU.}
\label{tab:time-sk}
\end{table}

\end{document}